\documentclass{ws-mpla}

\usepackage{epsf}

 \newcommand{\insertplot}[5]{\begin{figure}
 \hfill\hbox to 0.05in{\vbox to #5in{\vfill
 \inputplot{#1}{#4}{#5}}\hfill}
 \hfill\vspace{-.1in}
 \caption{#2}\label{#3}
 \end{figure}}
 \newcommand{\inputplot}[3]{
 \special{ps: plotfile #1}
}

\newcommand{\vphi}{\varphi}

\newcommand{\rd}{{\rm{d}}}

\begin{document}

\title{
Non-uniqueness, Counterrotation, and Negative Horizon Mass of
Einstein-Maxwell-Chern-Simons Black Holes}
 \vspace{1.5truecm}
\author{
{\bf Jutta Kunz and Francisco Navarro-L\'erida}
}
\address{
{
Institut f\"ur Physik, Universit\"at Oldenburg, Postfach 2503\\
D-26111 Oldenburg, Germany}
}
%\author{
%{\bf Francisco Navarro-L\'erida}
%}
%\affiliation{
%{Dept.~de F\'{\i}sica At\'omica, Molecular y Nuclear, Ciencias F\'{\i}sicas\\
%Universidad Complutense de Madrid, E-28040 Madrid, Spain}
%}
%\date{September 21, 2005}
\date{\today}

\maketitle

\begin{abstract}
Stationary black holes in 5-dimensional Einstein-Maxwell-Chern-Simons theory
possess surprising properties.
When considering the Chern-Simons coefficient $\lambda$ as a parameter,
two critical values of $\lambda$ appear: 
the supergravity value $\lambda_{\rm SG}=1$, and the value $\lambda=2$.
At $\lambda=1$,
supersymmetric black holes with vanishing horizon angular velocity,
but finite angular momentum exist.
As $\lambda$ increases beyond $\lambda_{\rm SG}$ 
a rotational instability arises,
and counterrotating black holes appear,
whose horizon rotates in the opposite sense to the angular momentum.
Thus supersymmetry is associated with the
borderline between stability and instability.
At $\lambda=2$ rotating black holes with vanishing angular momentum emerge.
Beyond $\lambda=2$ black holes may possess a negative horizon mass,
while their total mass is positive.
Charged rotating black holes with vanishing gyromagnetic ratio appear, and
black holes are no longer uniquely characterized by their global charges.
\end{abstract}

\ccode{PACS Nos.: 04.20.Jb, 04.40.Nr}

\section{Introduction}

Black holes in higher dimensions received much interest in recent years,
in particular in the context of string theory,
and with the advent of brane-world theories, 
raising the possibility of direct observation 
in future high energy colliders \cite{exp1,exp2}.

While the generalization of the Kerr metric to higher dimensions
was obtained long ago \cite{MP},
%by Myers and Perry \cite{MP} long ago,
the higher dimensional generalization of the Kerr-Newman solutions
of Einstein-Maxwell (EM) theory is still not known analytically \cite{KNP,KNV}.

EM black holes in $D$ dimensions are characterized by their mass $M$, charge $Q$, and 
$[(D-1)/2]$ angular momenta $\bf J$, their number corresponding to the
rank of the rotation group SO(D-1) \cite{MP}.
Their event horizon has surface gravity $\kappa$ and $(D-2)$-volume
$A_{\rm H}$, electrostatic potential $\Phi_{\rm H}$ and 
$[(D-1)/2]$ angular velocities ${\bf \Omega}_{\rm H}$.
These black holes satisfy the Smarr formula \cite{GMT}
\begin{equation}
M=  \frac{(D-2)}{(D-3)8\pi G_D} \kappa {A_{\rm H}}
+ \Phi_{\rm H} Q + \frac{(D-2)}{(D-3)}{\bf \Omega}_{\rm H} \cdot {\bf J} \ ,
\label{smarr}
\end{equation}
where $G_D$ is the $D$-dimensional Newton constant,
and the first law of black hole mechanics \cite{GMT},
\begin{equation}
\rd M= \frac{\kappa}{8\pi G_D} \rd {A_{\rm H}} + \Phi_{\rm H} \rd Q +
{\bf \Omega}_{\rm H} \cdot \rd {\bf J}\, .
\label{first}
\end{equation}

In odd dimensions, $D=2n+1$, the EM action
may be supplemented by a `$A\,F^n$' Chern-Simons (CS) term.
While not affecting the static black hole solutions,
this term significantly affects the properties
of stationary black hole solutions
\cite{BLMPSV,BMPV,GMT,surprise1,surprise2,KN1,KN2},
as discussed in this review.

The CS term also yields a modified Smarr formula,
where Eq.~(\ref{smarr}) is supplemented by a further term,
proportional to the CS coefficient $\lambda$ and
to the factor $(D-5)$ \cite{GMT}. Thus $D=5$ is a rather special case
among the class of odd-dimensional 
Einstein-Maxwell-Chern-Simons (EMCS) theories,
since the Smarr formula Eq.~(\ref{smarr}) remains unmodified.

The bosonic sector of minimal $D=5$ supergravity
may be viewed as a special case
of the general EMCS theory with Lagrangian
\begin{equation}
%{\cal L}= \frac{1}{16\pi G_5} \left[\sqrt{-g}(R -F^2) -
%\frac{2\lambda}{3\sqrt{3}}\varepsilon^{mnpqr}A_mF_{np}F_{qr}\right] \ ,
{\cal L}= \frac{1}{16\pi G_5} \sqrt{-g}\left[R -F^2 +
\frac{2\lambda}{3\sqrt{3}}\epsilon^{\mu \nu \rho \sigma \tau} 
A_\mu F_{\nu \rho}F_{\sigma \tau}\right] \ ,
\label{Lag}
\end{equation}
and CS coefficient $\lambda$, where $\lambda$ assumes 
the supergravity value $\lambda_{\rm SG}=1$.
When $\lambda=1$,
analytic solutions describing charged, rotating black holes
are known \cite{BLMPSV,BMPV,Cvetic1,Cvetic2,Cvetic3}.
In contrast, for $\lambda \ne 1$ charged rotating black hole solutions are known
only numerically \cite{KN1,KN2}.

The extremal limits of rotating charged black hole
solutions of $D=5$ EMCS theory with $\lambda=1$
are of special interest, since they encompass a
two parameter family of stationary supersymmetric black holes 
\cite{BLMPSV,BMPV}.
The mass of these supersymmetric black holes
is determined in terms of their charge
and saturates the bound \cite{Gibbons}
\begin{equation}
M \ge \frac{\sqrt{3}}{2} |Q| \ ,
\label{M-bound}
\end{equation}
while their two equal-magnitude angular momenta, $|J|=|J_1|=|J_2|$,
are finite and satisfy the bound \cite{BMPV,surprise2}
\begin{equation}
%\frac{|J|}{M^{3/2}} \le \frac{1}{2} \left( \frac{\sqrt{3} }{2} 
% \frac{Q}{M} \right)^{3/2}  \le \frac{1}{2} \ .
%|J| \le \frac{1}{2} \left( \frac{\sqrt{3} }{2} |Q| \right)^{3/2}
%J^2/M^3 \le 1/2 \ .
|J|^2 \le \frac{1}{6 \sqrt{3} \pi}  |Q|^3 \ ,
\label{J-bound}
\end{equation}
in units for which $G_5 =1$.
Their horizon angular velocities $\bf \Omega_{\rm H}$, however, vanish.
Thus their horizon is non-rotating, although their angular momentum
is nonzero. 
Clearly, angular momentum is stored in the Maxwell field,
but surprisingly, a negative fraction of the total angular momentum
is stored behind the horizon \cite{GMT,surprise2}.
The effect of rotation on the horizon is not to make it rotate
but to deform it into a squashed 3-sphere \cite{GMT}.

These special properties of $D=5$ supersymmetric EMCS black holes
caused speculations on how the properties
of $D=5$ black holes in general EMCS theories depend
on the CS coefficient \cite{GMT},
involving instability and counterrotation of black holes
beyond $\lambda=1$. Both these features turned out to be true \cite{KN1}.
Moreover further intriguing features appeared at and beyond the second
critical value $\lambda=2$,
involving non-uniqueness of black holes, the appearance of
non-static black holes with vanishing total angular momentum,
as well as the presence of black holes with negative horizon mass. 

\section{Properties of $5D$ EMCS Black Holes}

To obtain stationary EMCS black hole solutions,
we consider black hole space-times with bi-azimuthal symmetry,
implying the existence of three commuting Killing vectors,
$\xi = \partial_t$, $\eta_1=\partial_{\varphi_1}$, 
and $\eta_2=\partial_{\varphi_2}$
\cite{MP,frolov}.

While generic EMCS black holes possess two independent
angular momenta, we here restrict to black holes whose
angular momenta have equal magnitude.
The metric and the gauge field parametrization then
simplify.
In particular, for such equal-magnitude angular momenta
black holes, the general Einstein and gauge field equations reduce to
a set of ordinary differential equations \cite{KNP}, since the angular
dependence can be treated explicitly.

\subsection{Metric and gauge potential}

We employ a parametrization for the metric 
based on bi-azimuthal isotropic coordinates, 
well suited for the numerical construction 
\cite{KNP,KNV,KN1,KN2}
\begin{eqnarray}
  \rd s^2 = && -f \rd t^2
  +\frac{m}{f}\left( \rd r^2+r^2 \rd \theta^2 \right)
 +\frac{m-n}{f}\,r^2 \sin^2\theta \cos^2\theta  
 \left( \varepsilon_1 \rd \varphi_1 - \varepsilon_2 \rd \varphi_2 \right)^2 
 \nonumber \\ &&
  + \frac{n}{f}\, r^2 \left[\, \sin^2\theta
          \left( \varepsilon_1 \rd \varphi_1-\frac{\omega}{r} \rd t\right)^2
  +                            \cos^2\theta
          \left( \varepsilon_2 \rd \varphi_2-\frac{\omega}{r} \rd t\right)^2
 \right]
          \ , \label{metric} \end{eqnarray}
where $\varepsilon_k = \pm 1$, $k=1,2$, denotes the sense of rotation
in the $k$-th orthogonal plane of rotation.

The parametrization for the gauge potential, consistent with
Eq. (\ref{metric}), is
\begin{equation}
A_{\mu} \rd x^{\mu}=a_0 \rd t
+a_{\varphi} \left( \sin^2\theta \varepsilon_1 \rd \varphi_1
                  + \cos^2\theta \varepsilon_2 \rd \varphi_2 \right) \ .
\end{equation}
All metric and gauge field functions depend on $r$ only.

\subsection{Boundary conditions}

At infinity we impose on the metric the boundary conditions
$f=m=n=1$, $\omega=0$, 
i.e., the solutions are asymptotically flat.
For the gauge potential we choose a gauge, where $a_0=a_{\varphi}=0$.

The regular event horizon resides at a surface of constant radial coordinate,
$r=r_{\rm H}$ \cite{MP,KNP,KNV},
and is characterized by the condition $f(r_{\rm H})=0$ 
\cite{kkrot1,kkrot2,kkrot3,kkrot4,kkrot5,KNP,KNV}.
Here the metric functions satisfy the boundary conditions
$f=m=n=0$, $\omega=r_{\rm H} \Omega_{\rm H}$,
where $\Omega_{\rm H}$ is (related to) the horizon angular velocity,
defined in terms of the Killing vector
\begin{equation}
\chi = \xi + \Omega \left( \varepsilon_1 \eta_{1} 
                         + \varepsilon_2 \eta_{2} \right) \ ,
\label{chi} \end{equation}
which is null at the horizon. The gauge potential satisfies
$\left. \chi^\mu A_\mu \right|_{r=r_{\rm H}} =
\Phi_{\rm H} $,
%$ \left. \frac{\rd a_\vphi}{\rd r}\right|_{r=r_{\rm H}}=0$,
$ \left( \rd a_\vphi/\rd r \right)_{r=r_{\rm H}}=0$,
with constant horizon electrostatic potential $\Phi_{\rm H}$.

\subsection{Global properties}

Since the space-times we are considering are stationary,
bi-axisymmetric, and asymptotically flat we may compute the mass $M$ and
the two angular momenta $J_{(k)}$ of the black holes by means of the Komar expressions
associated with the respective Killing vector fields
\begin{equation}
M = \frac{-1}{16 \pi G_5} \frac{3}{2} \int_{S_{\infty}^{3}} \alpha \ , \ \ \
J_{(k)} = \frac{1}{16 \pi G_5}  \int_{S_{\infty}^{3}} \beta_{(k)} \ , \label{Komar_MJ}
\end{equation}
with $ \alpha_{\mu \nu \rho} =\epsilon_{\mu \nu \rho \sigma \tau}
\nabla^{\sigma} \xi^{\tau}$,
$\beta_{(k)\mu \nu \rho} =\epsilon_{\mu \nu \rho \sigma \tau}
\nabla^{\sigma} \eta_k^{\tau}$,
and for equal-magnitude angular momenta $J_{(k)}=\varepsilon_k J$,
$k=1$, 2. 

The electric charge $Q$ associated with the Maxwell field can be defined by
\begin{equation}
Q=\frac{-1}{8 \pi G_5} \int_{S_{\infty}^{3}} \tilde F \ , 
\label{elect_charge}\end{equation}
with ${\tilde F}_{\mu \nu \rho} \equiv  \epsilon_{\mu \nu \rho \sigma \tau} 
F^{\sigma \tau}$.

These global charges and the magnetic moment $\mu_{\rm mag}$,
can be obtained from the asymptotic expansions of the metric and the gauge
potential
\begin{equation}
f \rightarrow 1-\frac{8\, G_5 M }{3\pi r^2} \ , \ \ \
\omega \rightarrow \frac{4\, G_5 J }{\pi r^3} \ , \ \ \
a_0 \rightarrow  \frac{G_5 Q}{\pi r^2}  \ , \ \ \
a_{\varphi} \rightarrow - \frac{G_5 \mu_{\rm mag}}{ \pi r^2} 
\ . \end{equation}
The gyromagnetic ratio $g$ is then defined via
\begin{equation}
g=2\frac{M {\mu_{\rm mag}}}{Q J} \ . \label{g_factor}
\  \end{equation}

\subsection{Horizon properties}

The expansion at the horizon shows, that the surface gravity $\kappa$
\begin{equation}
\kappa^2 = -\frac{1}{2} \lim_{r \to r_{\rm H}} (\nabla_\mu \chi_\nu)
  (\nabla^\mu \chi^\nu) = \lim_{r \to r_{\rm H}} \frac{f}{(r-r_{\rm H}) \sqrt{m}}
  \   \label{surface_grav}
\end{equation}
is constant at the horizon,
as required by the zeroth law of black hole mechanics,
and that the electrostatic potential $\Phi_{\rm H}$
is constant at the horizon as well.
For equal-magnitude angular momenta,
the area of the horizon $A_{\rm H}$ reduces to
\begin{equation}
A_{\rm H}= 2 \pi^2 r_{\rm H}^{3}  \lim_{r \to r_{\rm H}}
\sqrt{\frac{m^{2} n}{f^{3}}} \ . \label{hor_area}
\end{equation}
To have a measure for the deformation of the horizon we consider
the circumferences of the horizon,
the polar circumference
$L_p$, where $\varphi_1$ and $\varphi_2$ are kept constant,
and the equatorial circumference,
$L_e$, where $\theta=0$ and $\varphi_1$ is kept constant
(or equivalently $\theta=\pi/2$ and $\varphi_2$ is kept constant).

The horizon mass $M_{\rm H}$ and horizon angular momenta
$J_{{\rm H} (k)}$ are given by
\begin{equation}
M_{\rm H} = \frac{-1}{16 \pi G_D} \frac{3}{2} \int_{{\cal H}} \alpha \ , \ \ \
J_{{\rm H} (k)} = \frac{1}{16 \pi G_D}  \int_{{\cal H}} \beta_{(k)} \ , \label{hor_MJ}
\end{equation}
where ${\cal H}$ represents the surface of the horizon,
and for equal-magnitude angular momenta
$J_{{\rm H} (k)} =\varepsilon_k J_{\rm H}$, $k=1$, 2.
The mass $M$ and angular momenta $J_{(k)}$ may thus be reexpressed in the
form \cite{GMT,surprise2}
\begin{equation}
 M=M_{\rm H} + M_{\Sigma} \ , \ \ \
 J_{(k)}=J_{{\rm H} (k)} + J_{\Sigma (k)} \ , \label{sum_MJ}
\end{equation}
where $J_{\Sigma (k)}$ is a `bulk' integral over the region of $\Sigma$
outside the horizon,
i.e.,
$\Sigma$ is a space-like hypersurface with boundaries at spatial
infinity and at the horizon.

These black holes satisfy the horizon mass formula
\begin{equation}
\frac{2}{3} M_{\rm  H} = \frac{\kappa A_{\rm H}}{8 \pi G_5} + 2 \Omega
J_{\rm H} \ , \label{hor_mass_form}
\end{equation}
and the Smarr formula \cite{GMT}
\begin{equation}
M = \frac{3}{2} \frac{\kappa A_{\rm H}}{8 \pi G_5} + \frac{3}{2} 2 \Omega
J  +  \Phi_{\rm H} Q \ . \label{smarr_like}
\end{equation}

\subsection{Scaling}

The system of ODEs is invariant under the scaling tranformation
\begin{equation}
r_{\rm H} \to \gamma r_{\rm H} \ , \ \ \Omega \to \Omega/\gamma \ , \ \
Q \to \gamma^{2} Q \ , \ \ 
a_\vphi \to \gamma a_\vphi \ , \label{scaling}
\end{equation}
with $\gamma>0$
\footnote{
Note, that for $D>5$ the CS coupling constant is dimensionful
and scales according to
${\lambda} \to \gamma^{n-2} {\lambda}$.
}.
The solutions then have the scaling symmetry
 $\tilde M= \gamma^2 M$,
 $\tilde J= \gamma^3 J$,
 $\tilde Q = \gamma^2 Q$,
 $\tilde r_{\rm H}=\gamma r_{\rm H}$,
 $\tilde \Omega = \Omega/\gamma$,
 $\tilde \kappa = \kappa/\gamma$, etc.

We also note, that
$\lambda \rightarrow - \lambda$ corresponds to $Q \rightarrow -Q$. Then,
without loss of generality we will assume $\lambda \ge 0$ hereafter.

\section{Numerical results}

For the numerical calculations, we introduce
the compactified radial variable $\bar{r}= 1-r_{\rm H}/r$ \cite{kkrot1,KNP}.
We employ a collocation method for boundary-value ordinary
differential equations, equipped with an adaptive mesh selection procedure
\cite{COLSYS}.
Typical mesh sizes include $10^3-10^4$ points.
The solutions have a relative accuracy of $10^{-10}$.
The set of numerical parameters to be fixed for a particular solution is
$\{r_{\rm H}, \Omega, Q, {\lambda}\}$.
By varying these parameters we generate families of EMCS black holes.

\subsection{Domain of existence}

Let us first consider the dependence of
the domain of existence of EMCS black holes
on the CS parameter $\lambda$.
In Fig.~1 we exhibit
the scaled angular momentum $|J|/M^{3/2}$ of extremal EMCS black holes 
versus the scaled charge $Q/M$ 
for several values of $\lambda$:
the pure EM case, $\lambda=0$,
the supergravity case, $\lambda=1$,
$\lambda=1.5$, the second critical case, $\lambda=2$,
and $\lambda=3$. 
For a given value of $\lambda$, 
black holes exist only in the regions bounded by the
$J=0$-axis and by the respective solid curves,
which are formed by extremal black holes.

For fixed finite $\lambda$,
there is an explicit asymmetry between solutions with
positive and negative electric charge.
The properties of EMCS black holes with
positive $Q$ are similar to those of EM black holes,
whereas for EMCS black holes with negative $Q$ surprising features are present.
\begin{figure}[t!]
\parbox{\textwidth}{
\centerline{
\mbox{
\hspace{0.5cm} {\small Fig.~1a} \hspace{-2.0cm}
\epsfysize=5.0cm \epsffile{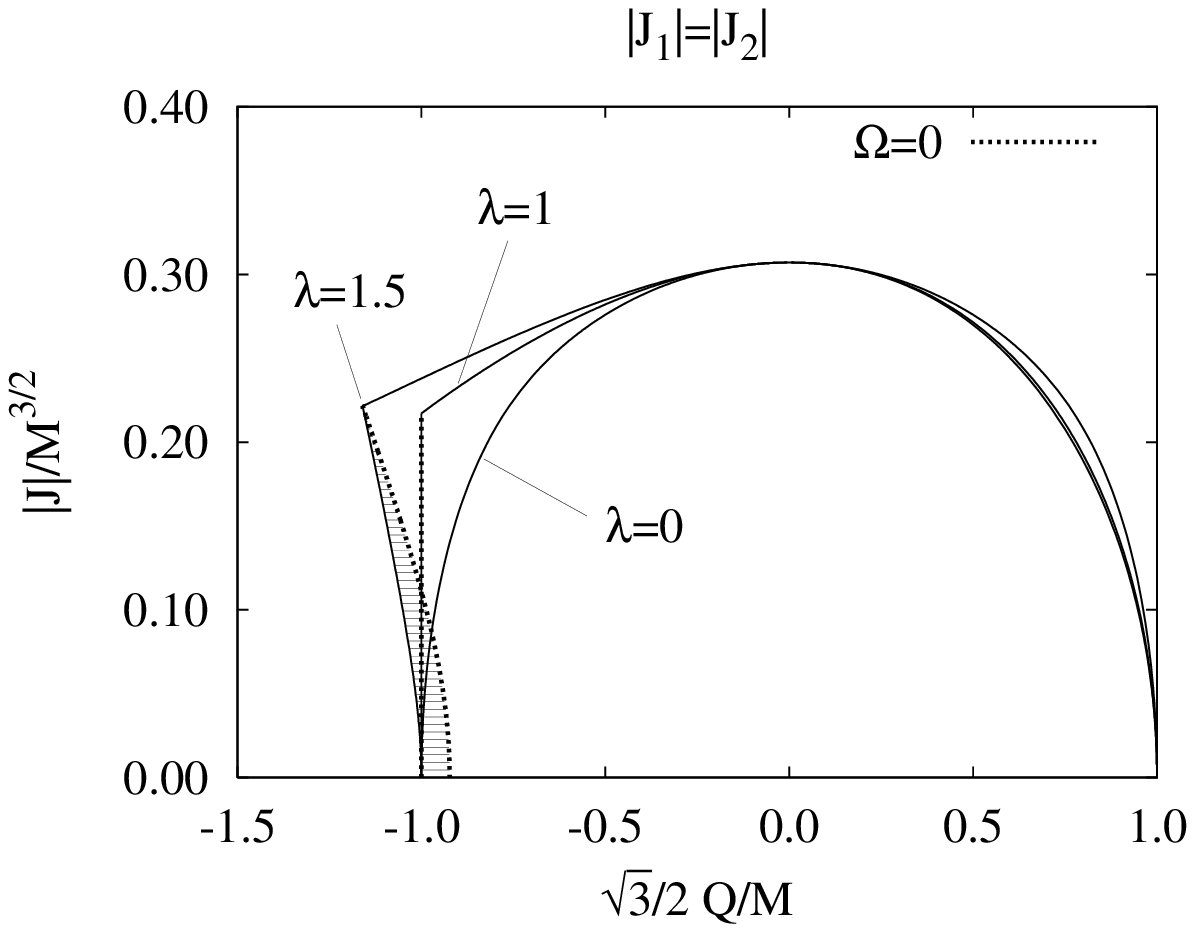} } \hspace{-1.cm}
\mbox{
\hspace{1.0cm} {\small Fig.~1b} \hspace{-2.0cm}
\epsfysize=5.0cm \epsffile{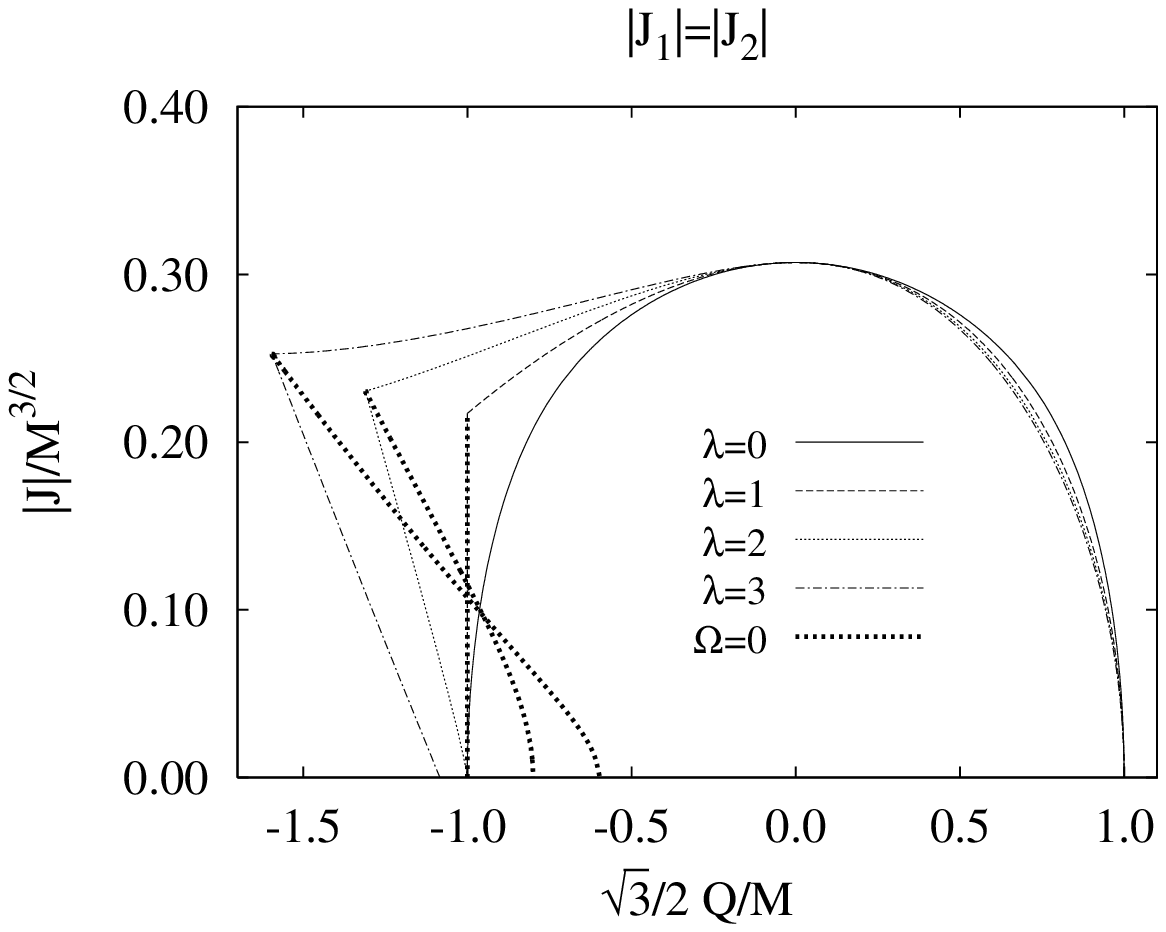} }
}}\vspace{0.2cm}
%{\bf Fig.~1} \small
\caption{
Scaled angular momentum $J/M^{3/2}$ versus scaled charge $Q/M$ for
extremal EMCS black holes and $\Omega=0$ solutions with CS coefficients
$\lambda=0$, 1, 1.5, 2, 3.
}
\vspace{0.0cm}
\end{figure}

As long as $\lambda<1$ the boundary is smooth,
but asymmetrical, unless $\lambda$ vanishes.
As $\lambda$ approaches the supergravity
value $\lambda=1$, the boundary develops a kink, 
which turns into a singular point, when $\lambda=1$.
In the supergravity case the boundary then consists of two parts:
a smooth curve and a straight vertical line,
where both parts join at the singular point.

When $1<\lambda \le 2$,
the static extremal black holes continue to reside at the lower 
left edge of the boundary of the domain of existence.
But beyond the second critical value, $\lambda=2$, 
they are located well within the domain of existence.
(Recall that the static black holes are independent of
$\lambda$.)

\subsection{$\Omega=0$ solutions}

The staticity theorem of EM theory claims, 
that stationary non-rotating black holes (i.e., 
black holes with vanishing horizon angular velocity)
are static \cite{staticity}.
EMCS black holes, in contrast, may possess a static horizon,
while their total angular momentum is non-vanishing \cite{BMPV}
\footnote{Black holes with static horizon
and finite total angular momentum
were first observed in Einstein-Maxwell-dilaton theory \cite{Rasheed}.}.

Not present below $\lambda=1$, these black holes arise
in EMCS theory at the critical value $\lambda_{\rm SG}$,
where they constitute the left vertical boundary of the 
domain of existence.
These $\lambda=1$ solutions are very special,
since they represent {\sl supersymmetric} charged rotating black holes,
where supersymmetry implies that the horizon must be non-rotating.
Since $\Omega=0$, these solutions possess no ergoregion.

These supersymmetric black holes saturate the mass bound Eq.~(\ref{M-bound}),
and they satisfy the angular momentum bound Eq.~(\ref{J-bound}).
Thus while angular momentum is built up,
the mass remains constant (for constant charge),
in accordance with the first law,
since $\kappa = \Omega =0$.

As the total angular momentum is increased from its static limiting
value $J=0$,
angular momentum is built up in the Maxwell field behind and outside
the horizon, as seen in Fig.~2a, where 
the total angular momentum $J$ and the horizon
angular momentum $J_{\rm H}$ are exhibited.
In particular, a negative fraction of the total angular momentum
is stored in the Maxwell field behind the horizon \cite{GMT}.
Thus, while one expects frame dragging effects 
to cause the horizon to rotate,
these effects are precisely counterbalanced by frame dragging effects,
due to the negative contribution to the angular momentum
by the fields behind the horizon,
allowing these black holes to retain a static horizon \cite{surprise2}.

All these supersymmetric black holes possess
a regular horizon, except for the limiting solution,
saturating the bound Eq.~(\ref{J-bound}). 
The area $A_{\rm H}$ of the horizon decreases as $|J|$ increases towards
the bound Eq.~(\ref{J-bound}),  
yielding a singular limiting solution 
with vanishing horizon area, $A_{\rm H}=0$.
The effect of the rotation on the horizon is not to make it rotate,
but to deform it from a round 3-sphere to a squashed 3-sphere
as seen in Fig.~2b \cite{GMT}. 

Along the non-supersymmetric branch $J$ and $J_{\rm H}$ have equal signs.

\begin{figure}[t!]
\parbox{\textwidth}{
\centerline{
\mbox{
\hspace{0.5cm} {\small Fig.~2a} \hspace{-2.0cm}
\epsfysize=5.0cm \epsffile{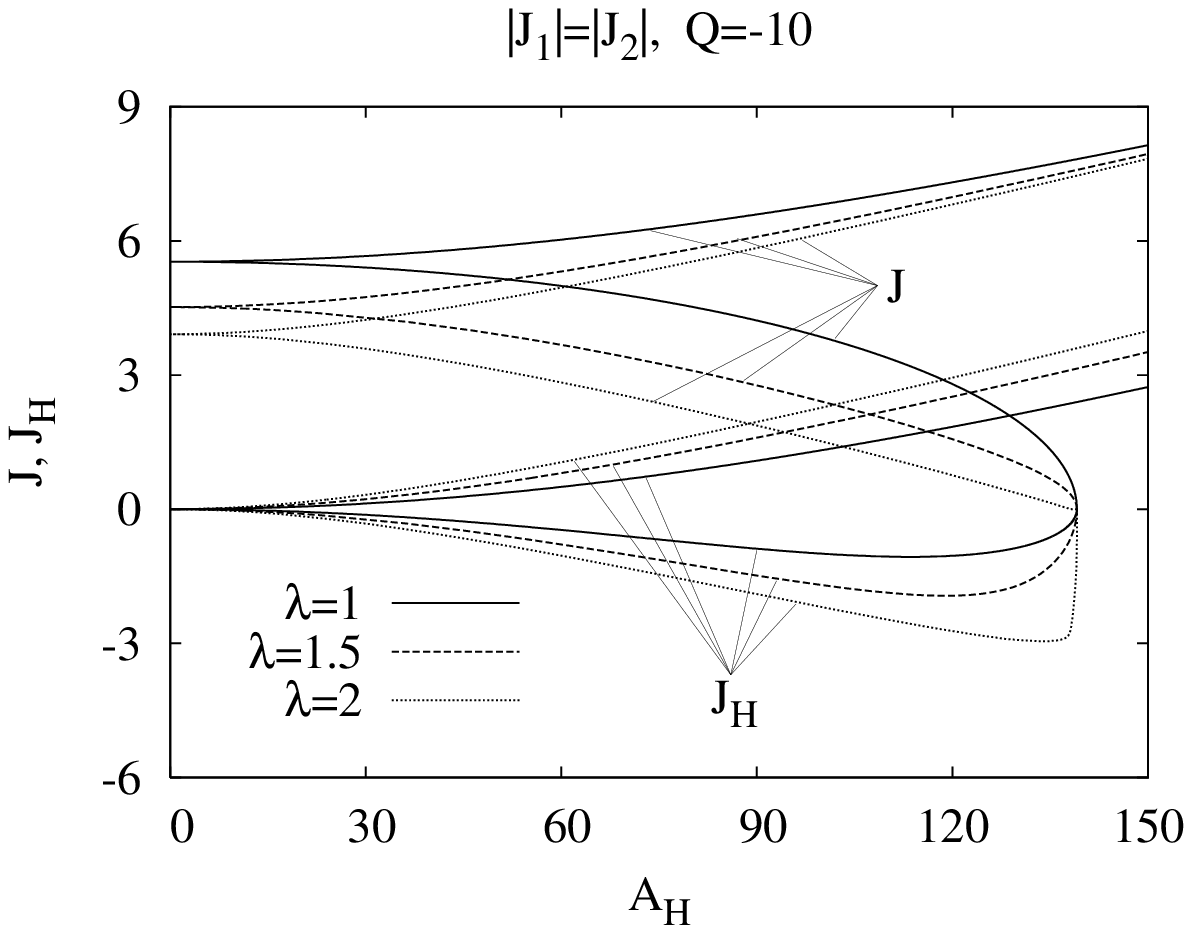} } \hspace{-1.cm}
\mbox{
\hspace{1.0cm} {\small Fig.~2b} \hspace{-2.0cm}
\epsfysize=5.0cm \epsffile{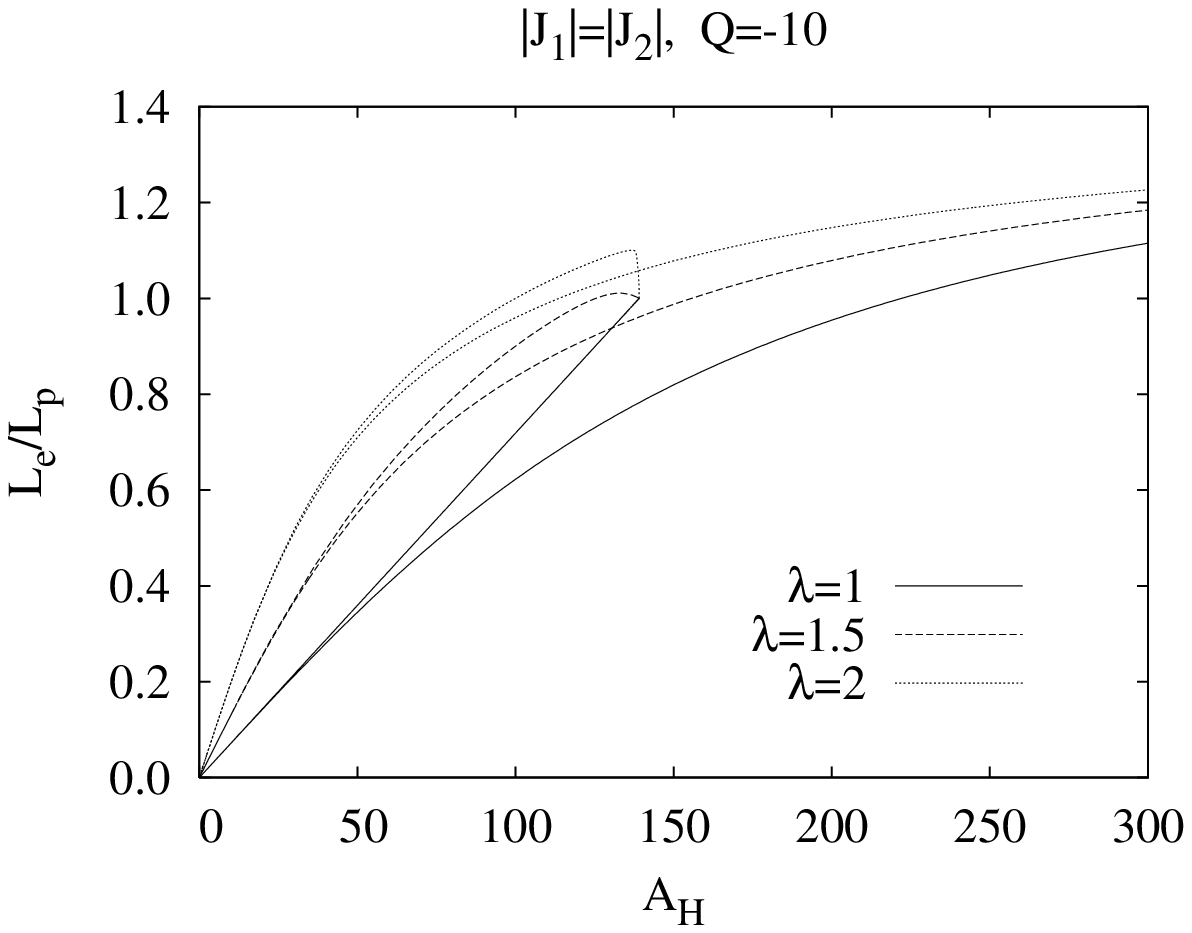} }
}}\vspace{0.2cm}
%{\bf Fig.~2} \small
\caption{
Properties of (almost) extremal black holes
($\lambda=1$, 1.5, 2; $Q=-10$).
a) Total angular momentum $J$ and horizon angular momentum $J_{\rm H}$,
b) deformation $L_{\rm e}/L_{\rm p}$, versus the area of the horizon $A_{\rm H}$.
}
\vspace{0.0cm}
\end{figure}

Below $\lambda=1$ the only $\Omega=0$ solutions are static black holes.
At $\lambda=1$, the supersymmetric black holes form a continuous set of 
non-static extremal $\Omega=0$ solutions.
As $\lambda$ is increased beyond its supergravity value, the
set of $\Omega=0$ solutions becomes non-extremal,
except for its extremal endpoint(s).
The location of such sets of $\Omega=0$ black holes 
within their respective domain of existence 
is indicated in Fig.~1
for several values of the CS coupling constant $\lambda>1$.

\subsection{Instability}

Extremal static EMCS black holes 
saturate the mass bound Eq.~(\ref{M-bound}) for any value of $\lambda$.
For stationary EMCS black holes demonstration of 
the bound Eq.~(\ref{M-bound}) relied on the fact
that $\lambda=1$ \cite{Gibbons,GMT}.
While the mass bound Eq.~(\ref{M-bound})
is respected by stationary black holes as long as $\lambda<1$,
it is violated, when $\lambda>1$, as seen in Fig.~1.

As speculated before \cite{GMT}, the mass can indeed
decrease with increasing angular momentum for fixed $\lambda>1$.
Thus while an extremal static black hole 
with zero Hawking temperature and spherical symmetry 
cannot decrease its mass by Hawking radiation,
it can however become unstable with respect to rotation, when $\lambda>1$,
with photons carrying away both energy and angular momentum
to infinity.
In terms of the first law as applied to $\lambda>1$
extremal black holes ($\kappa=0$) with fixed charge ($\rd Q=0$),
such an instability then requires, that 
the horizon is rotating in the opposite sense to the angular momentum,
since $\rd M = {\bf \Omega}_{\rm H} \cdot \rd {\bf J}$
must be negative.

In Fig.~3a we demonstrate explicitly,
that extremal static black holes can become
unstable with respect to rotation,
by exhibiting the deviation of the mass from the
static value, computed via the second-order derivative of the
mass with respect to the angular momentum at the extremal static solution.
For $1< \lambda < 2$ the mass is seen to decrease
with increasing magnitude of the angular momentum for fixed electric charge,
as attributed to the presence of counterrotating solutions,
connected to the extremal static solution.
Thus supersymmetry marks indeed the
borderline between stability and instability.

\begin{figure}[t!]
\parbox{\textwidth}{
\centerline{
\mbox{
\hspace{0.5cm} {\small Fig.~3a} \hspace{-2.0cm}
\epsfysize=5.0cm \epsffile{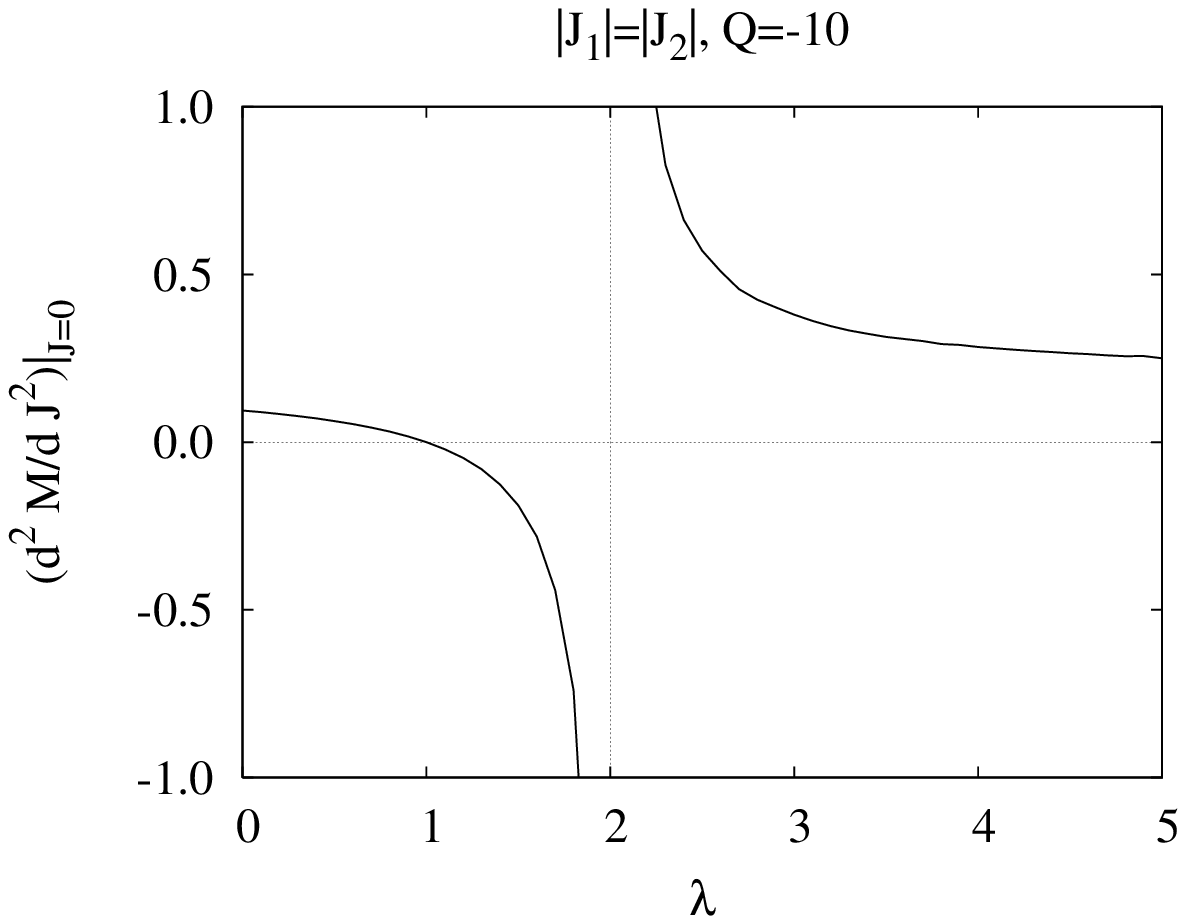} } \hspace{-1.cm}
\mbox{
\hspace{1.0cm} {\small Fig.~3b} \hspace{-2.0cm}
\epsfysize=5.0cm \epsffile{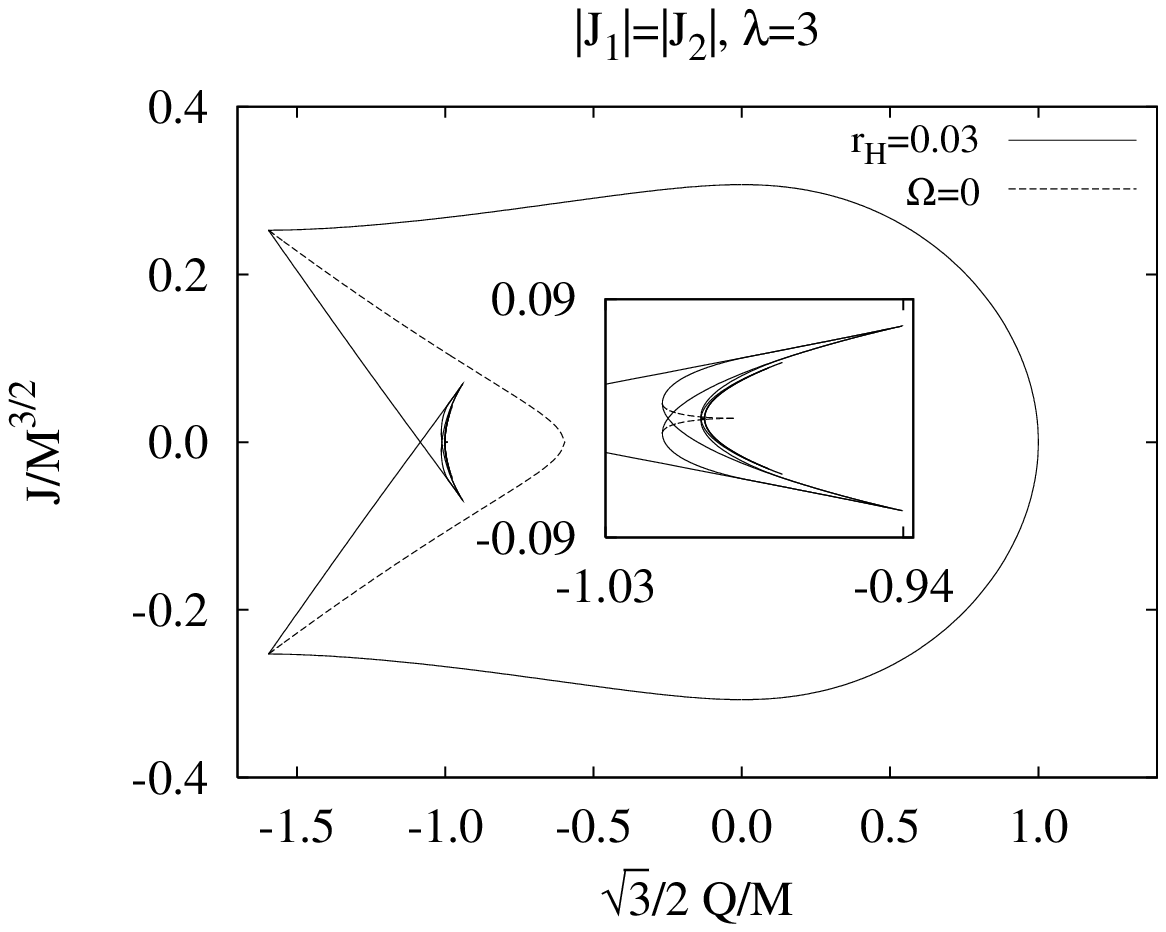} }
}}\vspace{0.2cm}
%{\bf Fig.~3} \small
\caption{
a) Second-order derivative of the mass $M$ with respect to the angular momentum
$J$ at $J=0$ versus the CS coupling constant $\lambda$ for extremal
black holes ($Q=-10$).
b)
Scaled angular momentum $J/M^{3/2}$ versus scaled charge $Q/M$
for (almost) extremal and $\Omega=0$ solutions with CS coefficient $\lambda=3$.
}
\vspace{0.0cm}
\end{figure}

\subsection{Counterrotation}

Responsible for the onset of instability, counterrotating black holes
appear beyond the critical value $\lambda=1$. 
For a given value of $\lambda$,
the (main) set of rotating $\Omega=0$ black holes
(extending from a static solution to the corresponding
extremal singular solution)
then divides the domain of existence into two parts, as seen in Fig.~1.
The right part contains ordinary black holes,
where the horizon rotates in the same sense as the angular momentum,
whereas the left part (the shaded region in Fig.~1a) 
contains black holes with unusual properties.
When $1<\lambda<2$, all black holes in this region are counterrotating
\footnote{
Also counterrotating black holes were first observed in
Einstein-Maxwell-dilaton theory \cite{KKN-c}.},
i.e., their horizon rotates in the opposite sense to the angular momentum 
\cite{KN1,KN2}.
When $\lambda \ge 2$ black holes with further intriguing features appear.

\subsection{$J=0$ black holes}

As expected from the change in stability,
another special case is reached, when $\lambda=2$.
Indeed, as $\lambda$ reaches $2$, another new phenomenon arises:
a (continuous) set of rotating $J=0$ solutions appears 
and persists as $\lambda$ is increased beyond $2$
\cite{strau}.
The existence of these rotating $J=0$ solutions
relies on a special partition of the total
angular momentum $J$, where the angular momentum within the horizon 
$J_{\rm H}$ is precisely equal and opposite to 
the angular momentum in the Maxwell field outside the horizon.
In contrast, for $\lambda<2$ only static $J=0$ solutions exist.
The presence of $J=0$ solutions is seen in Figs.~3b and 4a for $\lambda=3$.

\begin{figure}[t!]
\parbox{\textwidth}{
\centerline{
\mbox{
\hspace{0.5cm} {\small Fig.~4a} \hspace{-2.0cm}
\epsfysize=5.0cm \epsffile{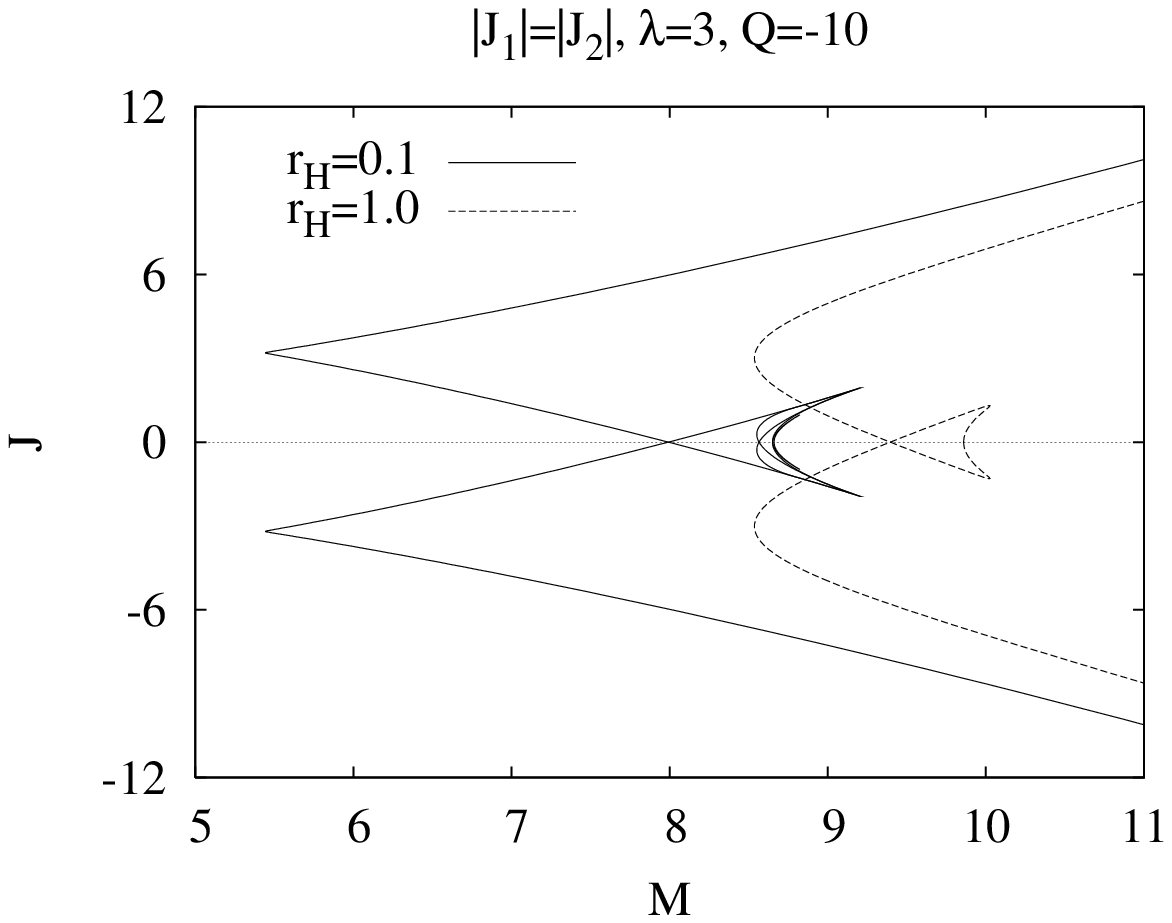} } \hspace{-1.cm}
\mbox{
\hspace{1.0cm} {\small Fig.~4b} \hspace{-2.0cm}
\epsfysize=5.0cm \epsffile{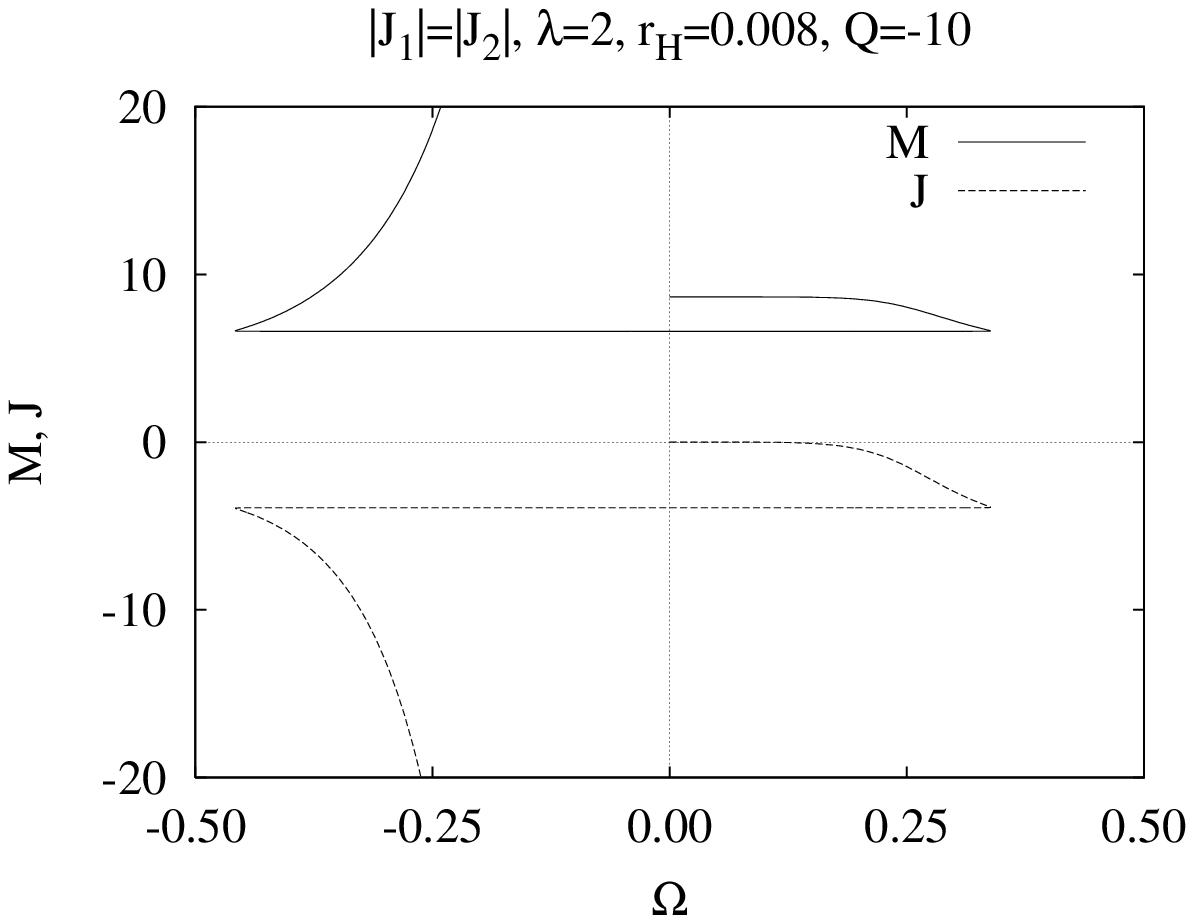} }
}}\vspace{0.2cm}
%{\bf Fig.~4} \small
\caption{
a)
Angular momentum $J$ versus mass $M$
for non-extremal black holes with horizon radii $r_{\rm H}=0.1$ and 1.0
($\lambda=3$; $Q=-10$).
b) Angular momentum $J$ and mass $M$
versus horizon angular velocity $\Omega$ for almost extremal black holes
($\lambda=2$, $r_{\rm H}=0.008$, $Q=-10$).
}
\vspace{0.0cm}
\end{figure}

Related to the $J=0$ solutions, the domain of existence of EMCS black holes 
changes character, and extremal rotating $J=0$ solutions replace
extremal static solutions as the left edge of the boundary 
on the $J=0$ axis, when $\lambda>2$. 

Thus beyond $\lambda=2$ the set of extremal solutions
not only forms the boundary of the domain of existence,
but continues well within this domain,
until the static extremal black hole is reached
in a complicated pattern of bifurcating branches.
Insight into this set is gained in Fig.~3b, where (almost)
extremal solutions are exhibited for CS coefficient $\lambda=3$,
together with non-static $\Omega=0$ solutions.
Note, that all this new structure arises well within
the counterrotating region, in the vicinity of the static extremal black holes.

With increasing $\lambda$ an increasing number of such extremal
rotating $J=0$ black holes appears, and with them
an increasing number of extremal $J \ne 0$ solutions
with non-rotating horizon, each forming the end point
of a whole set of (non-extremal)
$\Omega=0$, $J \ne 0$ black holes.
Indeed, these sets of $\Omega=0$ solutions always appear in pairs, 
connecting a static solution with an extremal solution,
whose mass assumes an extremal value as well,
in accordance with the first law.

The numerical data further indicate, that at the critical value $\lambda=2$
a (continuous) set of extremal rotating $J=0$ black holes with constant mass 
is present. This is illustrated in Fig.~4b.
As the horizon angular velocity $\Omega$ increases,
their mass $M$ can remain constant, as long as $J=0$,
and the angular momentum is redistributed appropriately
(as indicated by the steep rise of $J_{\rm H}$ in Fig.~2a).
For these black holes, the deformation is oblate.

\subsection{Non-uniqueness}

Fig.~4a further reveals that beyond $\lambda=2$ 
black holes are no longer uniquely characterized by their global charges.
Thus the uniqueness conjecture does not hold for $D=5$ EMCS
stationary black holes with horizons of spherical topology,
provided $\lambda>2$
\footnote{
The previous counterexamples involved black rings 
\cite{blackrings1,blackrings2,blackrings3,blackrings4,blackrings5}.}.
For $\lambda=2$ even an infinite set of extremal
black holes with the same global charges appears to exist,
as numerical data indicate (see Fig.~4b).

\subsection{Four types of black holes}

To explore the properties of $\lambda>2$ EMCS black holes further,
let us now consider non-extremal black holes.
We exhibit in Fig.~5 a set of solutions for $\lambda=3$,
possessing constant charge $Q=-10$ and
constant (isotropic) horizon radius $r_{\rm H}=0.2$.
Fig.~5a and 5b show the total angular momentum $J$
and the horizon angular momentum $J_{\rm H}$, respectively,
as functions of the horizon angular velocity $\Omega$,
Fig.~5c and 5d show the corresponding masses $M$ and $M_{\rm H}$,
and Fig.~5e and 5f the gyromagnetic ratio $g$ and the deformation 
of the horizon as measured by the ratio of equatorial and polar circumferences
$L_{\rm e}/L_{\rm p}$.

\begin{figure}[t]
\parbox{\textwidth}{
\centerline{
\mbox{
\hspace{0.5cm} {\small Fig.~5a} \hspace{-2.0cm}
\epsfysize=5.0cm \epsffile{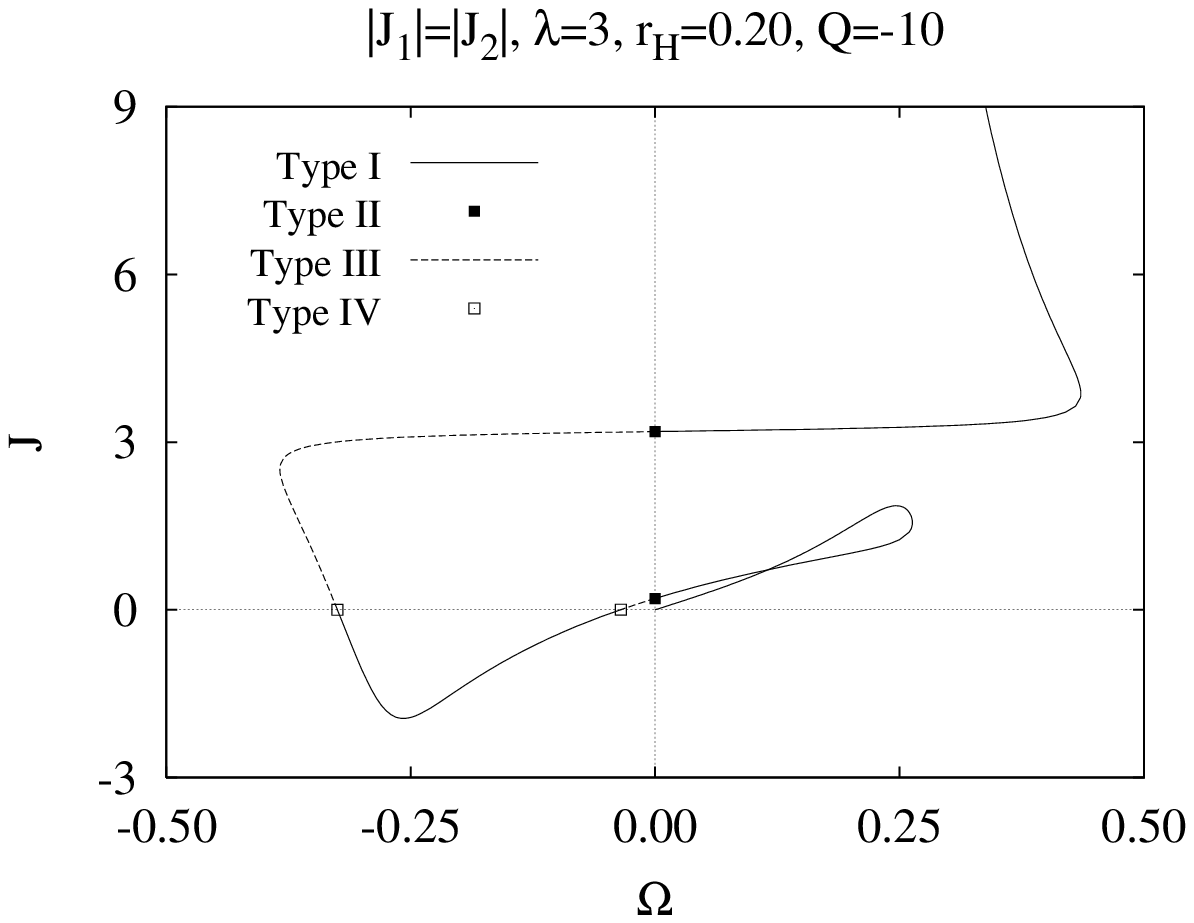} } \hspace{-1.cm}
\mbox{
\hspace{1.0cm} {\small Fig.~5b} \hspace{-2.0cm}
\epsfysize=5.0cm \epsffile{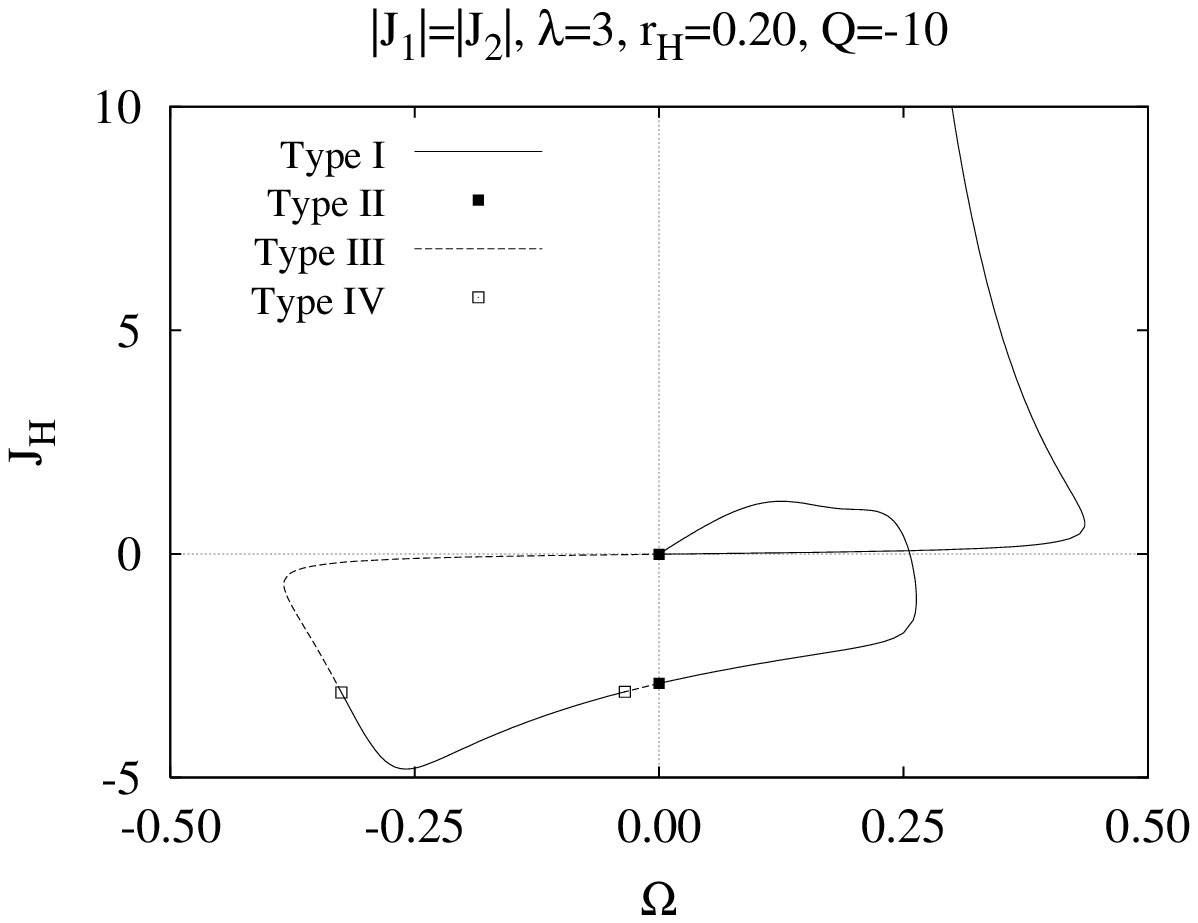} }
}\vspace{0.cm} }
\parbox{\textwidth}{
\centerline{
\mbox{
\hspace{0.5cm} {\small Fig.~5c} \hspace{-2.0cm}
\epsfysize=5.0cm \epsffile{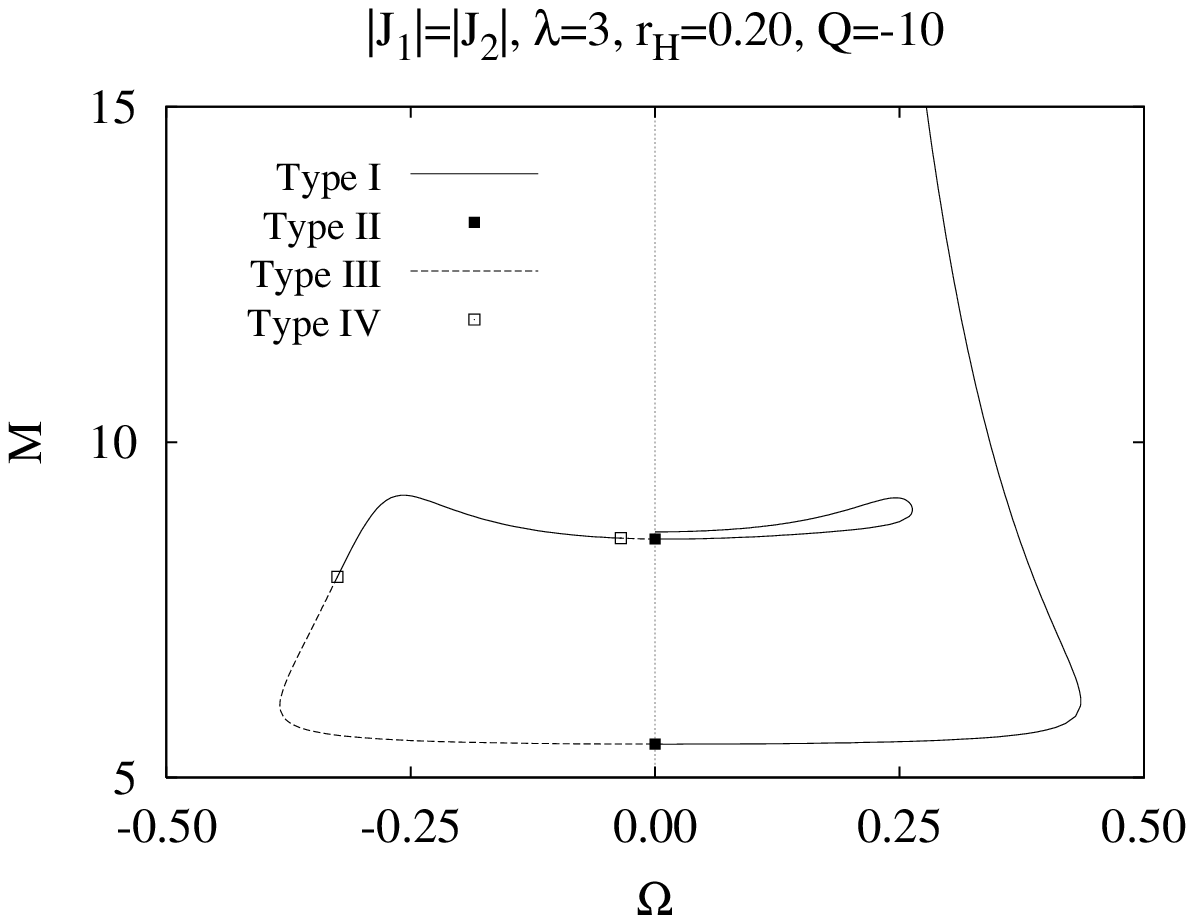} } \hspace{-1.cm}
\mbox{
\hspace{1.0cm} {\small Fig.~5d} \hspace{-2.0cm}
\epsfysize=5.0cm \epsffile{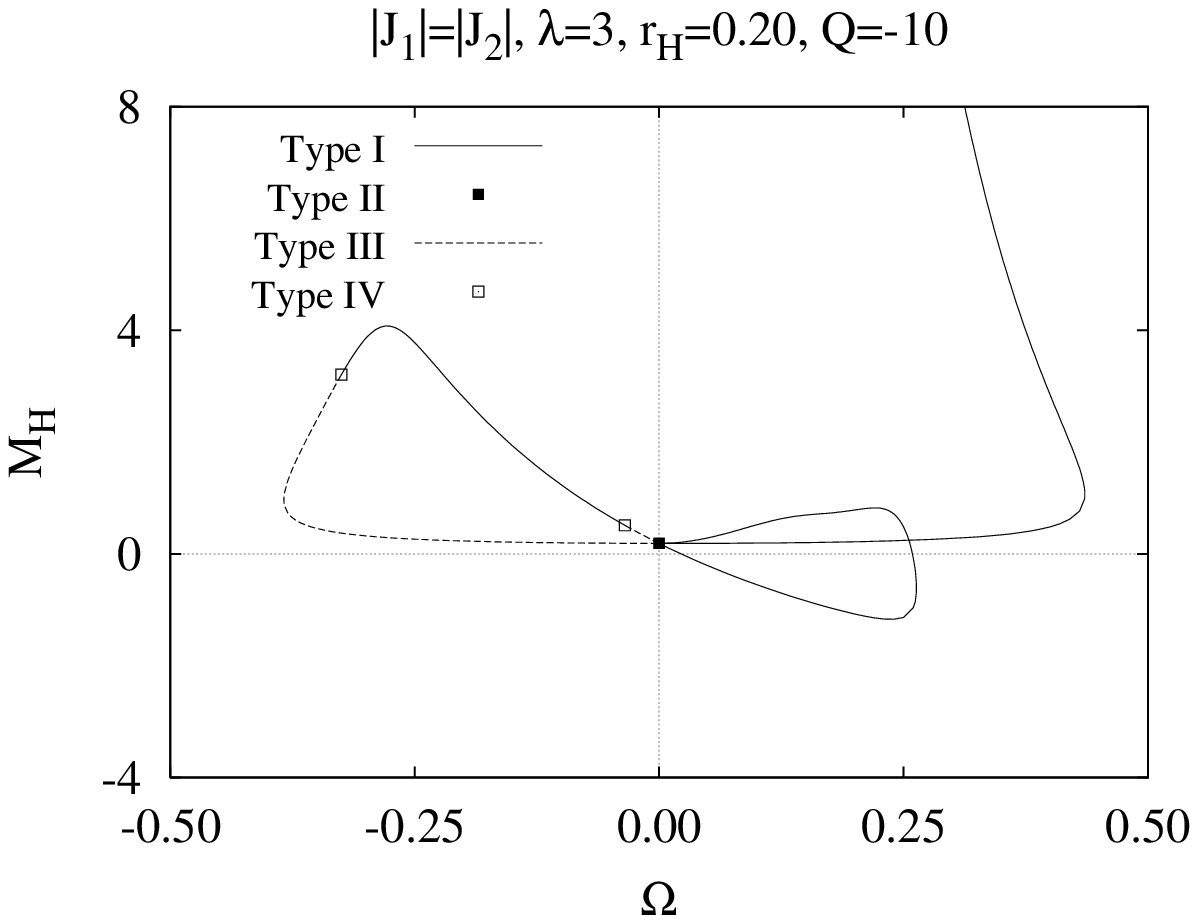} }
}\vspace{0.cm} }
\parbox{\textwidth}{
\centerline{
\mbox{
\hspace{0.5cm} {\small Fig.~5e} \hspace{-2.0cm}
\epsfysize=5.0cm \epsffile{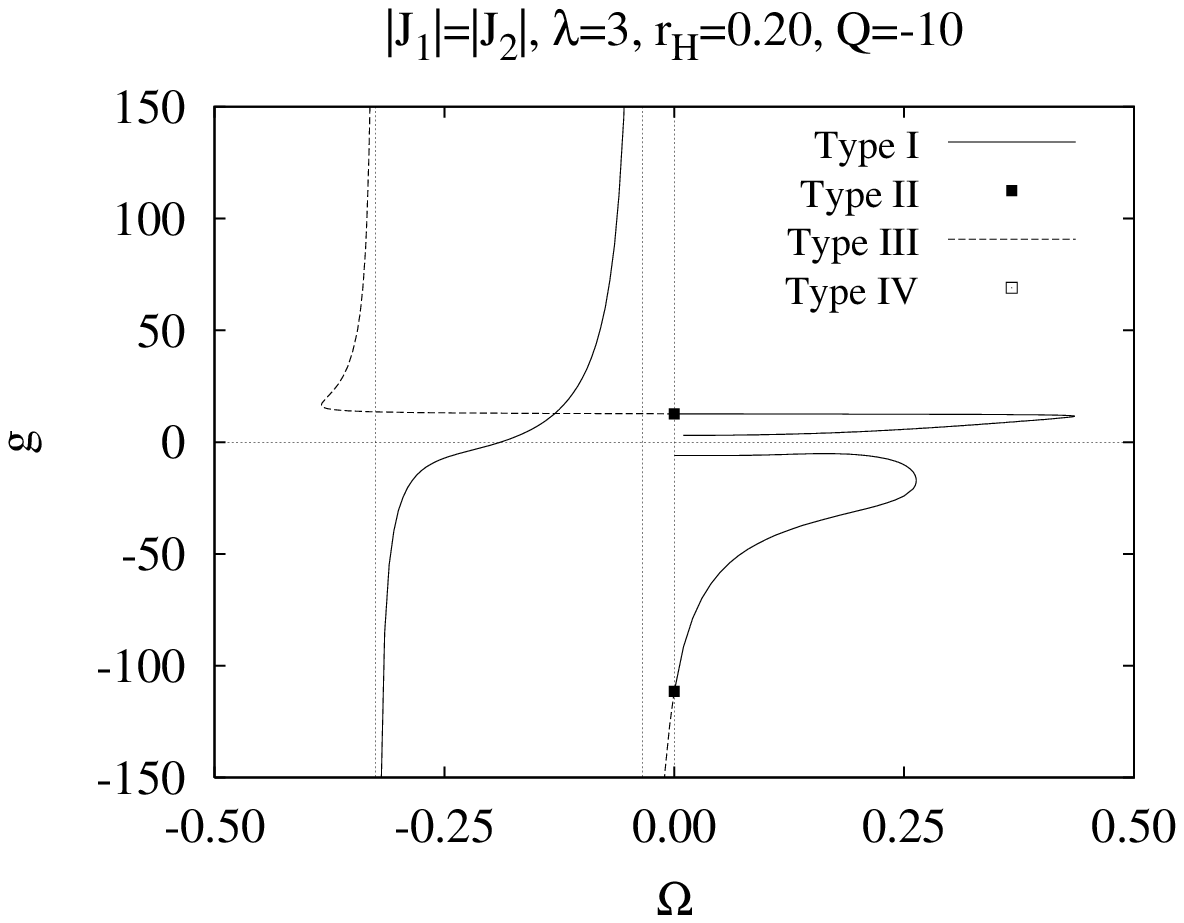} } \hspace{-1.cm}
\mbox{
\hspace{1.0cm} {\small Fig.~5f} \hspace{-2.0cm}
\epsfysize=5.0cm \epsffile{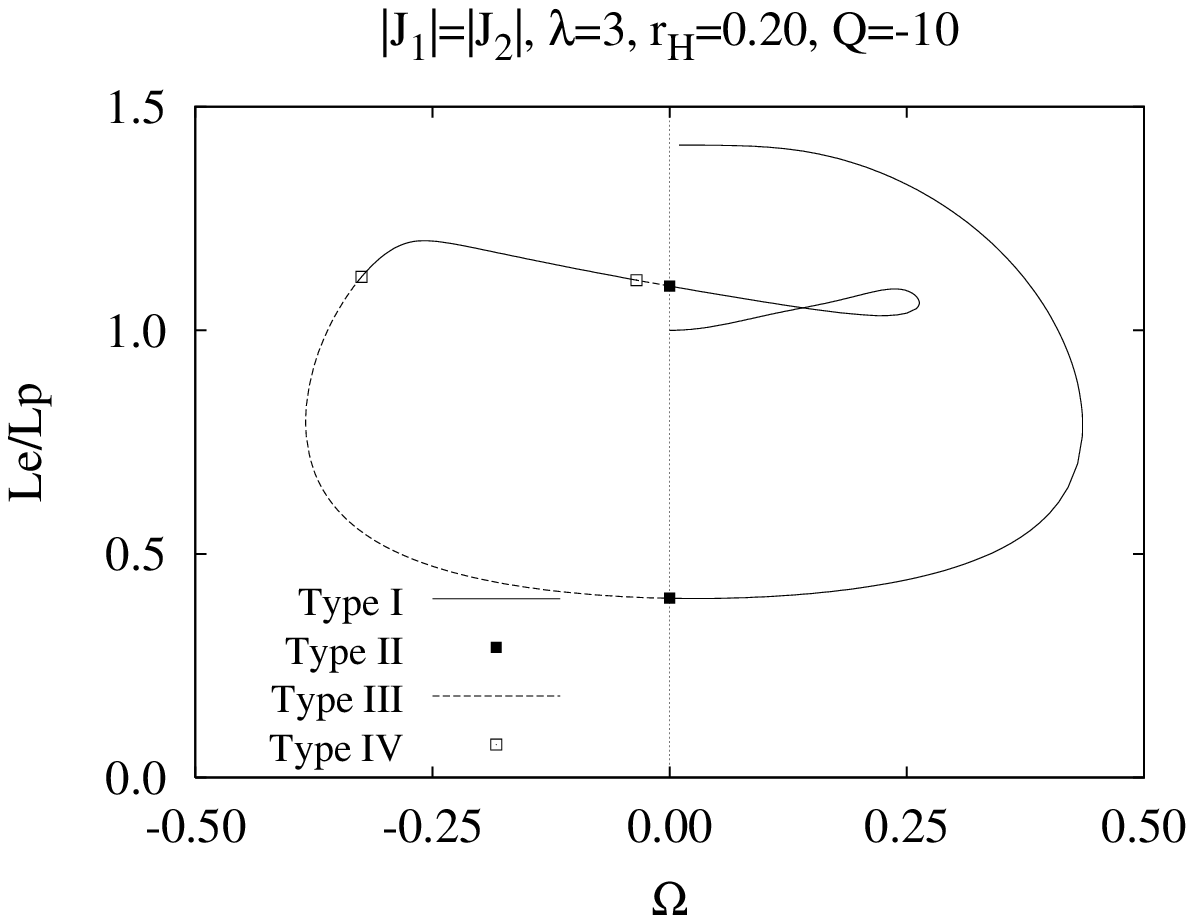} }
}}\vspace{0.2cm}
\caption{
%{\bf Fig.~5} \small
Properties of non-extremal $\lambda=3$ EMCS black holes with
charge $Q=-10$ and horizon radius $r_{\rm H}=0.2$.
a) Angular momentum $J$, b) horizon angular momentum $J_{\rm H}$,
c) mass $M$, d) horizon mass $M_{\rm H}$,
e) gyromagnetic ratio $g$, and
f) ratio of horizon circumferences $L_{\rm e}/L_{\rm p}$
%f) horizon area $A_{\rm H}$
versus horizon angular velocity $\Omega$.}
\vspace{0.0cm}
\end{figure}

Fig.~5a exhibits the four types of rotating black holes,
as classified by their total angular momentum $J$
and horizon angular velocity $\Omega$: Type I black
holes correspond to the corotating regime, i.e., $\Omega J \ge 0$, and
$\Omega=0$ if and only if $J=0$ (static). Type II black holes possess a static horizon
($\Omega=0$), although their angular momentum does not vanish ($J\neq 0$).
Type III black holes are characterized by counterrotation,
i.e., the horizon angular velocity
and the total angular momentum have oposite signs, $\Omega J < 0$.
Type IV black holes, finally, possess a rotating horizon ($\Omega \neq
0$) but vanishing total angular momentum ($J=0$).

\subsection{Negative horizon mass}

As seen in Fig.~5b, the horizon angular momentum $J_{\rm H}$ of these
black holes need not have the same sign as the total angular momentum $J$,
and neither does the `bulk' angular momentum $J_{\Sigma}$.
As the horizon of the black hole is set into rotation,
angular momentum is stored in the Maxwell field both behind
and outside the horizon, yielding a rich variety of configurations.
Starting from the static solution, a corotating branch evolves,
along which $J_{\rm H}$ and $J_\Sigma$ have opposite signs.
After the first bifurcation $\Omega$ moves back towards zero
and so does $J$, but both $J_{\rm H}$ and $J_\Sigma$ remain finite,
retaining part of their built up angular momentum
and thus their memory of the path, like in a hysteresis.
This is important, since when moving $\Omega$ continuously
back to and beyond zero,
the total angular momentum follows and changes sign as well.
The horizon angular momentum, however, retains its sign.
Thus the product $\Omega J_{\rm H}$ turns negative
and remains negative up to the next bifurcation
and still further, until $\Omega$ reaches again zero.

The presence of $\Omega J_{\rm H}<0$ solutions
explains the occurrence of black holes
with negative horizon mass, $M_{\rm H}<0$, 
which are exhibited in Fig.~5d.
The correlation between $\Omega J_{\rm H}<0$ and $M_{\rm H}<0$ black holes
is evident here. In fact,
the sets of $\Omega J_{\rm H}<0$ and $M_{\rm H}<0$ black holes almost coincide,
when $\kappa A_{\rm H}$ is small,
as seen from the horizon mass formula Eq.~(\ref{hor_mass_form}).
The angular momentum stored in the Maxwell field behind the horizon
is thus responsible for the negative horizon mass of the black holes.
The total mass is always positive, however, as seen in Fig.~5c.

\begin{figure}[t!]
\parbox{\textwidth}{
\centerline{
\mbox{
\hspace{0.5cm} {\small Fig.~6a} \hspace{-2.0cm}
\epsfysize=5.0cm \epsffile{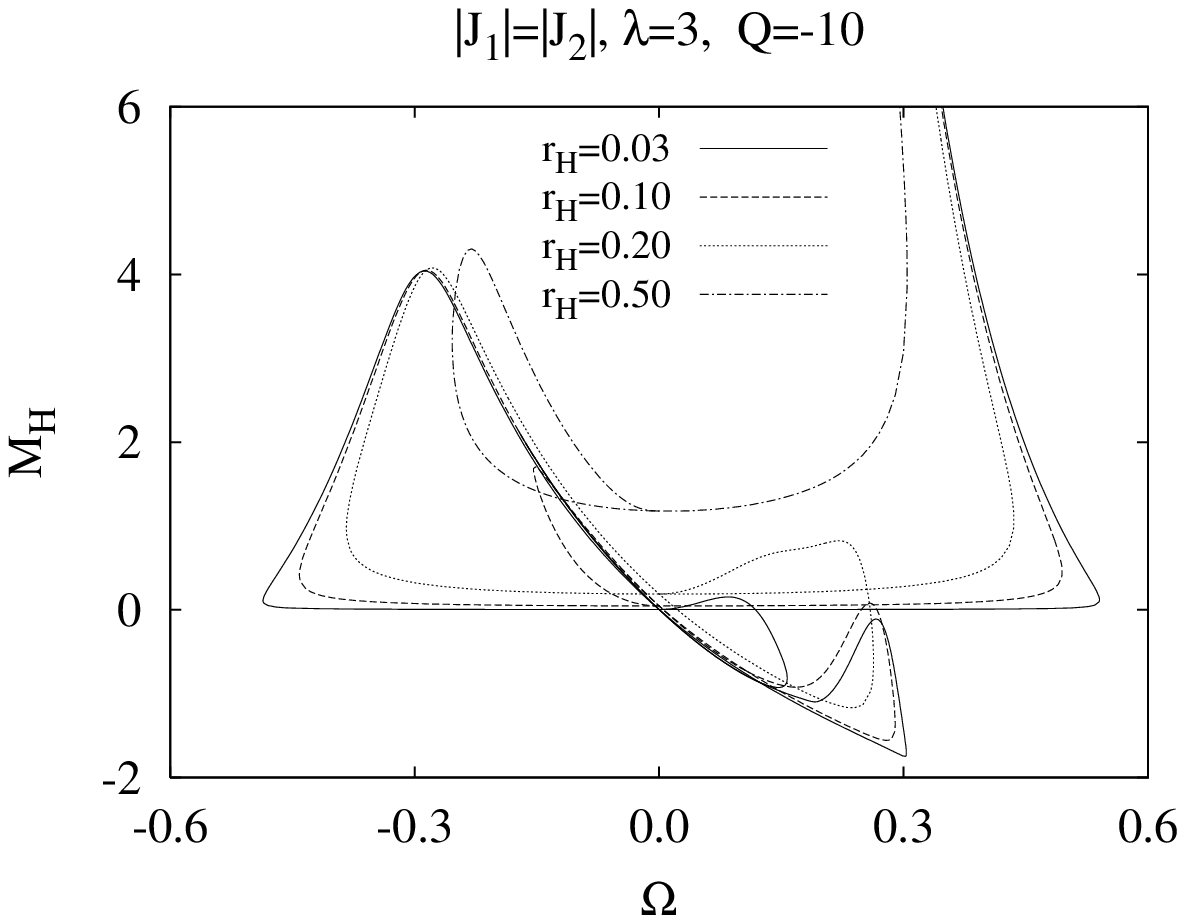} } \hspace{-1.cm}
\mbox{
\hspace{1.0cm} {\small Fig.~6b} \hspace{-2.0cm}
\epsfysize=5.0cm \epsffile{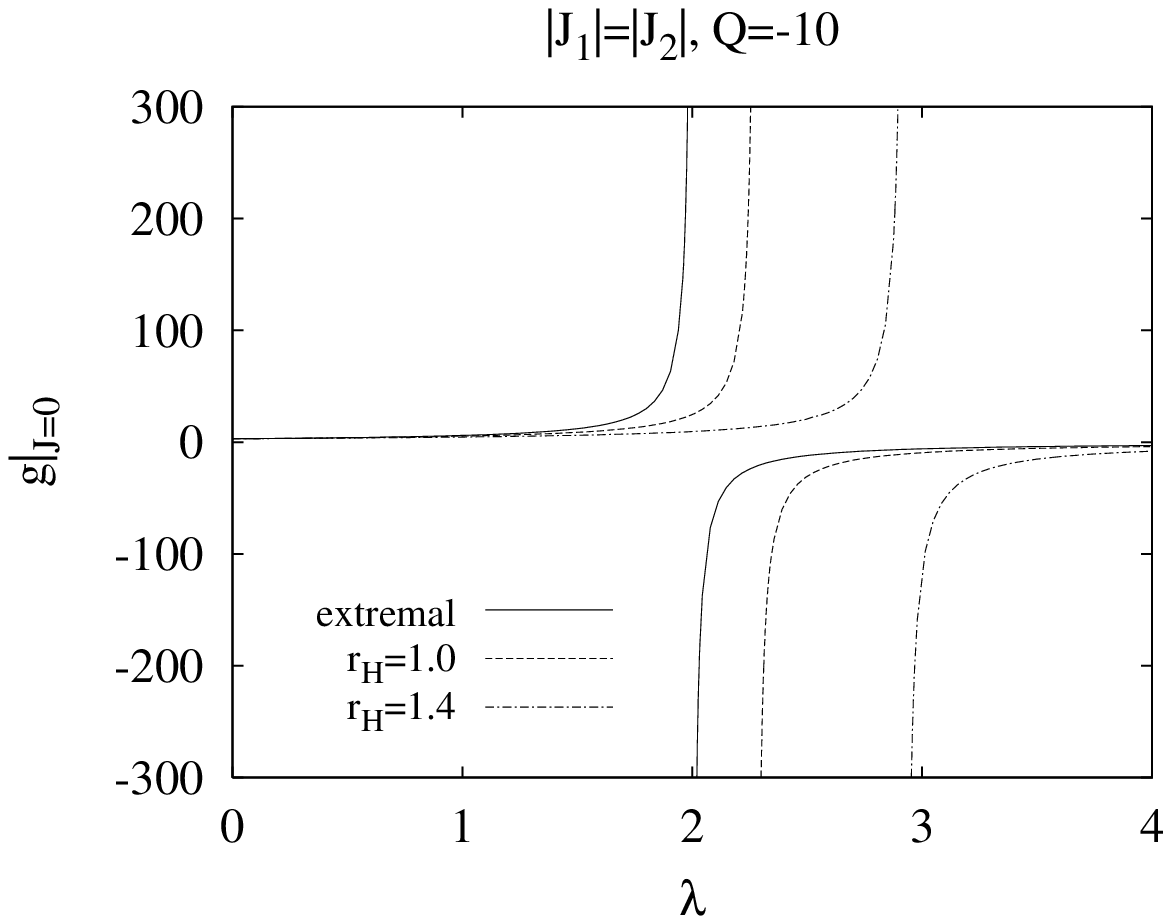} }
}}\vspace{0.2cm}
%{\bf Fig.~2} \small
\caption{
a) Horizon mass $M_{\rm H}$ versus horizon angular velocity $\Omega$
for black holes with horizon radii $r_{\rm H}=0.03$, 0.1, 0.2 and 0.5
($\lambda=3$; $Q=-10$).
b) Limiting value for $J \to 0$ of the gyromagnetic ratio $g$ versus the CS coupling constant $\lambda$ for
black holes with $r_{\rm H}=0$, 1, 1.4 ($Q=-10$).
}
\vspace{0.0cm}
\end{figure}

For larger values of $r_{\rm H}$ the set of negative horizon
mass black holes decreases, 
while it increases for smaller values of $r_{\rm H}$, as seen in Fig.~6a.
In fact, as $r_{\rm H}$ decreases, more branches of solutions appear
in the vicinity of the static solution, giving rise to more branches
of negative horizon mass black holes.

\subsection{Gyromagnetic ratio}

Another interesting feature of these charged rotating EMCS black holes
is their gyromagnetic ratio $g$, exhibited in Fig.~5e.
When $\lambda > 2$,
the gyromagnetic ratio is unbounded, reaching any real value
including zero.
The main consequence of this is that, contrary to pure EM theory,
a vanishing total angular momentum does not readily
imply a vanishing magnetic moment and viceversa.

The gyromagnetic ratio is thus another indicator of the
particularity of $\lambda = 2$, as also seen in Fig.~6b.

\section{Conclusions}

Stationary black holes of EM theory in 4 dimensions
now appear to be very special, since
their familiar properties are not shared in general by black holes
in higher dimensions or by black holes in the presence
of more or different fields.

\subsection{$D=5$ EMCS black holes}

Charged rotating black holes
of EMCS theory in five dimensions,
which are asymptotically flat 
with a regular horizon of spherical topology,
can exhibit remarkable features. 
Classifying the EMCS black holes by their total angular momentum $J$
and horizon angular velocity $\Omega$, four types of black holes arise:
I. corotating black holes, i.e., $\Omega J \ge 0$, for all values of $\lambda$,
II. black holes with static horizon and
non-vanishing total angular momentum, i.e., $\Omega=0$, $J\neq 0$,
for $\lambda \ge 1$,
III. counterrotating black holes, where the horizon angular velocity
and the total angular momentum have opposite signs, i.e., $\Omega J < 0$,
for $\lambda > 1$,
and IV. black holes with rotating horizon and
vanishing total angular momentum, i.e., $\Omega \neq 0$, $J=0$,
for $\lambda \ge 2$.

As the horizon of static EMCS black holes is set into rotation,
angular momentum is stored in the Maxwell field both behind
and outside the horizon.
Following paths through configuration space,
the horizon angular momentum $J_{\rm H}$
and the `bulk' angular momentum $J_\Sigma$
can retain part of the angular momentum built up in the Maxwell field,
even when the horizon angular velocity vanishes again.
Thus they retain the memory of the path, like a hysteresis.
Consequently, solutions with $\Omega J_{\rm H}<0$
appear, which possess a negative horizon mass $M_{\rm H}<0$,
as long as $\kappa A_{\rm H}$ is sufficiently small.
Thus the angular momentum stored in the Maxwell field behind the horizon
is responsible for the negative horizon mass of these black holes.
Their total mass is positive, however \footnote{
The occurrence of a negative horizon mass has
been also reported recently in the context of $4D$ black holes
surrounded by perfect fluid rings \cite{ansorg}.}.

Moreover, these EMCS black holes are no longer uniquely characterized
by their global charges,
i.e., the uniqueness conjecture does not hold in general
for stationary black holes with horizons of spherical topology;
and their gyromagnetic ratio
may take any real value, including zero.

\subsection{$D=4$ EMD black holes}

A number of features of 
$D=5$ EMCS black holes appear also for $D=4$ EMD black holes,
with dilaton coupling constant $\gamma$,
when both electric ($Q$) and magnetic ($P$) charge are present \cite{KKN-c}.
In EMD theory the Kaluza-Klein value $\gamma=\sqrt{3}$
represents the (first) critical value.
For $\gamma<\sqrt{3}$ only corotating black holes exist.
For $\gamma=\sqrt{3}$ stationary black holes with non-rotating horizon
appear and form the vertical part of the boundary \cite{Rasheed}.
Their angular momentum is bounded by $|J| \le |PQ|$ 
(analogous to Eq.~(\ref{J-bound})).
For $\gamma>\sqrt{3}$ corotating and counterrotating black holes 
exist \cite{KKN-c}.
Thus in EMD theory the Kaluza-Klein value $\gamma=\sqrt{3}$ marks 
the change from stability to instability.
Stationary $\Omega=0$ black holes and counterrotating
black holes also exhibit squashed horizons \cite{KKN-c}.

Whether also $J=0$, $\Omega\ne 0$ solutions are present,
whether uniqueness is violated, or whether negative horizon
mass black holes are present remains to be investigated.

\subsection{$D>5$ EMCS black holes}

All these intriguing new types of stationary $5D$ EMCS black holes
occur also for EMCS black holes
in higher odd dimensions \cite{KN2}. But since
the CS coefficient then becomes dimensionful,
it changes under scaling transformations Eq.~(\ref{scaling}), unless
${\lambda=0}$.
Thus any feature present for a certain
non-vanishing value of ${\lambda}$
will be present for any other non-vanishing value
(although for the correspondingly scaled value of the charge).
For that reason, the critical behaviour of the solutions
cannot be classified by ${\lambda}$ alone, but only by a scale invariant
ratio involving ${\lambda}$ \cite{KN2}.

Interestingly, in $D=9$, a further type of black holes appears:
V. non-static $\Omega=J=0$ black holes,
possessing a finite magnetic moment \cite{KN2}.
Clearly, further surprises may be waiting in higher dimensions.

\section*{Acknowledgments}

FNL gratefully acknowledges Ministerio de Educaci\'on y Ciencia for
support under grant EX2005-0078.

%\section*{References}

\end{document}